\documentclass[runningheads]{llncs}
\usepackage[T1]{fontenc}
\usepackage{graphicx}
\usepackage{amsmath}
\usepackage{tabularray} 
\usepackage{subcaption}
\usepackage{booktabs} 
\usepackage{xcolor, soul} 
\usepackage{array}
\usepackage[section]{placeins}
\begin{document}
\title{Constraint-Based Model in Multimodal Learning to Improve Ventricular Arrhythmia Prediction}
\titlerunning{Constraint-Based Multimodal Learning}
%



\author{Evariste Njomgue Fotso\inst{1} \and
Buntheng Ly\inst{2} \and
Hubert Cochet\inst{2} \and \\ Maxime Sermesant\inst{1,2}}

\authorrunning{Evariste Njomgue Fotso et al.}

\institute{Inria, Université Côte d'Azur, Epione team, Sophia Antipolis, France \and
IHU Liryc, Université de Bordeaux, Bordeaux, France\\
\email{Contact: maxime.sermesant@inria.fr}}

\maketitle              
\begin{abstract}
Cardiac disease evaluation depends on multiple diagnostic modalities: electrocardiogram (ECG) to diagnose abnormal heart rhythms, and imaging modalities such as Magnetic Resonance Imaging (MRI), Computed Tomography (CT) and echocardiography to detect signs of structural abnormalities.
Each of these modalities brings complementary information for a better diagnosis of cardiac dysfunction. 
However, training a machine learning (ML) model with data from multiple modalities is a challenging task, as it increases the dimension space, while keeping constant the number of samples. In fact, as the dimension of the input space increases, the volume of data required for accurate generalisation grows exponentially.
In this work, we address this issue, for the application of Ventricular Arrhythmia (VA) prediction, based on the combined clinical and CT imaging features, where we constrained the learning process on medical images (CT) based on the prior knowledge acquired from clinical data.
The VA classifier is fed with features extracted from a 3D myocardium thickness map (TM) of the left ventricle. The TM is generated by our pipeline from the imaging input and a Graph Convolutional Network is used as the feature extractor of the 3D TM.
We introduce a novel Sequential Fusion method and evaluate its performance against traditional Early Fusion techniques and single-modality models.
The cross-validation results show that the Sequential Fusion model achieved the highest average scores of $80.7 \% \pm 4.4$ Sensitivity and $73.1 \% \pm 6.0$ F1 score, outperforming the Early Fusion model at $65.0 \% \pm 8.9$ Sensitivity and $63.1 \% \pm 6.3$ F1 score.
Both fusion models achieved better scores than the single-modality models, where the average Sensitivity and F1 score are $62.8 \% \pm 10.1; 52.1 \% \pm 6.5$ for the clinical data modality and $62.9 \% \pm 6.3; 60.7 \% \pm 5.3$ for the medical images modality.

%

\end{abstract}
\section{Introduction}
Sudden Cardiac Death (SCD), when the heart stops beating suddenly, is recognised as a high-priority public health topic. SCD is the most common cause of death worldwide, accounting for 4.25 million deaths every year \cite{aer_journal_neil}. 
Among  causes of SCD, ischaemic heart disease (also called coronary heart disease) is the most common cause \cite{aer_journal_neil}.
Ischaemic heart disease is a significant risk factor for the development of Ventricular Arrhythmia, 
such as Ventricular Tachycardia (VT) and Ventricular Fibrillation (VF).
Despite intensive research on the topic, over the past 25 years, the absolute number of Cardiovascular disease (CVD)
cases has increased in Europe and in the EU, with increases in the number of new CVD cases found in most countries \cite{eu_heart_network}. 
The most significant advance in the prevention of SCD has been the development of the implantable cardioverter-defibrillator (ICD) \cite{ref_icd_nanthakumar}.On the one hand people are still dying of SCD and on the other hand, the current  prescription guidelines of ICDs which is left ventricular ejection fraction (LVEF) <30–35\% \cite{Russo_icd} capture only a mere 20\% all SCD \cite{Wellens_scds}.

To enhance early risk identification of SCD, Dakun Lai et al. have proposed a machine learning model on measurable arrhythmic markers derived from ECG signals \cite{Dakun_scd_pred}. O’Mahony et al., have presented an individualised risk estimates for SCD in hypertrophied cardiomyopathy patients, considering various clinical parameters \cite{Mahony_scd_pred}. Global Electric Heterogeneity (GEH) parameters, quantifying the abnormal electrophysiological substrate, have also been identified as independent factors associated with SCD risk, improving risk prediction when combined with other clinical characteristics \cite{Jonathan_scd_pred}. For paediatric patients, the HCM Risk-Kids model estimates the risk of SCD and highlights the importance of considering both clinical and genetic factors in risk assessment \cite{Norrish_scd_pred}.
In order to improve the current clinical criteria for ICD candidacy, a combination (fusion) of clinical characteristics with other markers may significantly improve risk stratification \cite{Nikolaos_markers}, but feature fusion is still a challenging research topic in machine learning \cite{survey_fusion}.
In the literature, feature concatenation is the primary technique used for data fusion in multimodal classification \cite{William_multimodal_review}. This concatenation is typically performed either at the beginning of the learning process (Early Fusion) or at the end of the learning process (Late Fusion).


In this study, we propose a novel fusion technique for multimodal VA classification (positive VA+ or negative VA-). The technique can also handle the class imbalance issue which is particularly prevalent in medical diagnosis, where the occurrence of positive cases (VA+) is significantly rarer than negative cases. Our Sequential Fusion technique progressively constrains the learning process on higher dimensional data (medical images modality), based on knowledge acquired on lower dimensional data (clinical data). The knowledge acquired on clinical data, called prior modality, is 
used in the loss function of the learning process on medical images modality. We also study the contribution of each modality in our fusion technique.

\section{Method}

\subsection{Sequential Fusion Model}

Inspired by constraints-based models, we propose the Sequential Fusion model. A constraints-based model refers to a type of modelling approach that incorporates constraints to guide the learning process or restrict the possible solutions. These constraints can be mathematical expressions, logical statements, or any other form of limitation that helps narrow down the solution space and ensure that the model produces results that adhere to certain predefined criteria.

\begin{figure}[!ht]
  \centering
  \includegraphics[width=1\linewidth]{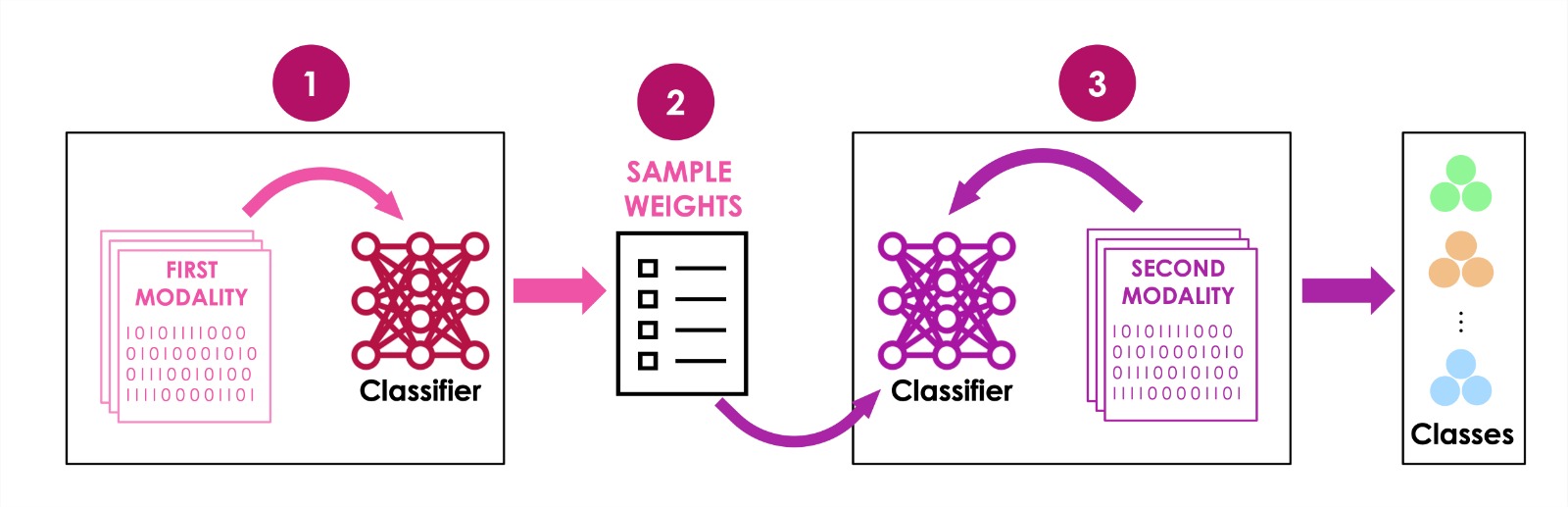}
  \caption{Sequential Fusion Model: general setting.}
  \label{fig:seq_fusion_model}
\end{figure}

To setup a Sequential Fusion model with two modalities (fig. \ref{fig:seq_fusion_model}), a first classifier is trained only with the first modality (the prior modality). The classification result on the prior modality is used to compute the prior modality constraint as defined in the next chapter. This constraint is then added into the learning process with the second modality to get the final output.

\subsection{Prior Modality Constraint Formulation}

The first model is trained using only clinical data which is the prior modality: it provides prior knowledge for the final classifier on imaging data from the classification errors on the clinical data. We account for this prior knowledge with the following weighting strategy.

Let us consider two sets $s_1$ and $s_2$, where $s_1$ is the set of well-classified samples and $s_2$ the set of misclassified samples using the prior modality. The prior modality weights ($prior\_weight_i$) is computed using the eq. \ref{eq:prior_weights}, where $prior\_weight_i$ is the weight for each set $s_i$,  $n$ is the total number of samples, $k$ is the number of sets (two in our setting) and $n_{s_i}$ is the number of samples in each set $s_i$.
%
\begin{equation}\label{eq:prior_weights}
prior\_weight_i = \frac{n}{k*n_{s_i}}
\end{equation}
The prior modality weights ($prior\_weight_i$) are the mathematical formulation of the prior-knowledge, where more importance is given to misclassified samples by the prior modality so that the final classifier is more focused on these samples. 

Moreover, in our setting, the data set is unbalanced. To tackle the unbalanced class issue, we use the same samples weighting technique, to give higher weight to minority class (VA-) and lower weight to the majority class (VA+).

To control both of these effects, we define two hyper-parameters $\alpha, \beta \in [0, 1]^2$ and two weighting strategies: same weighting, defined in eq. \ref{eq:same_strategy}, and stratified weighting, defined in eq. \ref{eq:strat_strategy}.


\begin{equation}\label{eq:same_strategy}
w_{l} = \left\{
    \begin{array}{l}
        \alpha * class\_weight_j + \beta * prior\_weight_i \\
    \end{array}
\right.
\end{equation}

\begin{equation}\label{eq:strat_strategy}
w_{l} = \left\{
    \begin{array}{ll}
        \alpha * class\_weight_j & \mbox{for $s_1 (i=1)$} \\
        \alpha * class\_weight_j + \beta * prior\_weight_i & \mbox{for $s_2 (i=2)$}
    \end{array}
\right.
\end{equation}

In equations \ref{eq:same_strategy} and \ref{eq:strat_strategy}, $class\_weight_j$ corresponds to weights which correct the imbalance effect. The weight of each class $j$ ($class\_weight_j$) is computed by the same formulation in eq. \ref{eq:prior_weights}, where $n$ is the total number of samples, $k$ is the number of classes (two in our setting) and $n_{c_j}$ is the number of samples in each class $c_j$. In eq. \ref{eq:prior_weights}, $n_{s_i}$ is replaced by $n_{c_j}$ for the computation of $class\_weight_j$.


In the same weighting strategy (eq. \ref{eq:same_strategy}),  equal importance is given to both the effects of unbalanced class and misclassified samples with the prior modality, whereas in stratified weighting strategy (eq. \ref{eq:strat_strategy}) more importance is given to positive (VA+) class and misclassified samples by the prior modality.

The loss function (binary cross-entropy) of the learning process on the medical images modality is then defined in eq. \ref{eq:loss_im_modality}, where $y_l$ is the true label of the $l$-th sample (1 for the positive class VA+ and 0 for the negative class VA- and $p_l$ is the predicted probability of the $l$-th sample belonging to the positive class.

\begin{equation}\label{eq:loss_im_modality}
Loss_{BCE} = w_l*y_llog(p_l) + w_l*(1 - y_l)log(1 - p_l)
\end{equation}

The Early Fusion model is trained with both weighting strategies and only the best one in terms of our performance metrics is kept (F1 score and Sensitivity). The novelty of our approach lies in these weights ($prior\_weight_i$), which is computed based on the knowledge acquired from the prior modality. This formulation can be generalised in multi-class classification with any additional constraints required in the learning process.

\section{Materials}

\subsection{Data set and Features Extraction}


We used a retrospective data set of myocardial infarction patients collected at a hospital between 2010 and 2020.
Patients with history of surgical procedure on the LV were excluded from the cohort.
In this work, patient's cardiac CT image, with the scan date more than 1 month after the infarction, and clinical descriptions were collected.
We classified as VA+ the patients, who have experienced sustained VT, VF and abort cardiac arrest.
Finally, we obtained a study data set of 600 patients, with 165 and 435 of VA+ and VA- patients, respectively. Table ~\ref{features_details_tab} gives details on available clinical data modality.

\begin{table}[h]
    \centering
    \caption{Clinical Data Modality (N=600)}
    \label{features_details_tab}
    \begin{tabular}{lc}
        \toprule
        Feature Name & Statistics: [min, max] mean or proportion \\
        \midrule
        LVEF \% & [10, 78] 45.07$\pm 13.36$ \\
        Age (years) & [27, 99] 72.69$\pm 12.09$ \\
        Smoking & 270 (45\%) \\
        Dyslipidaemia & 436 (73\%) \\
        Diabete & 159 (27\%) \\
        Hypertension & 394 (66\%) \\
        Sex & 100 Females / 500 Males \\
        VA & Positive (VA+) 165 (28\%) \\
        \midrule
        \multicolumn{2}{p{\linewidth}}{\textbf{Description:} LVEF \% - Left Ventricular Ejection Fraction, Age - Patient age at the scan date.} \\
    \end{tabular}
\end{table}


To extract the thickness map features from CT images, we used the automatic pipeline proposed by a cardiac imaging group in 2022. \cite{buntheng_ref}.

\subsection{Experimental setup}

\begin{figure}[!ht]
  \centering
  \includegraphics[width=1\linewidth]{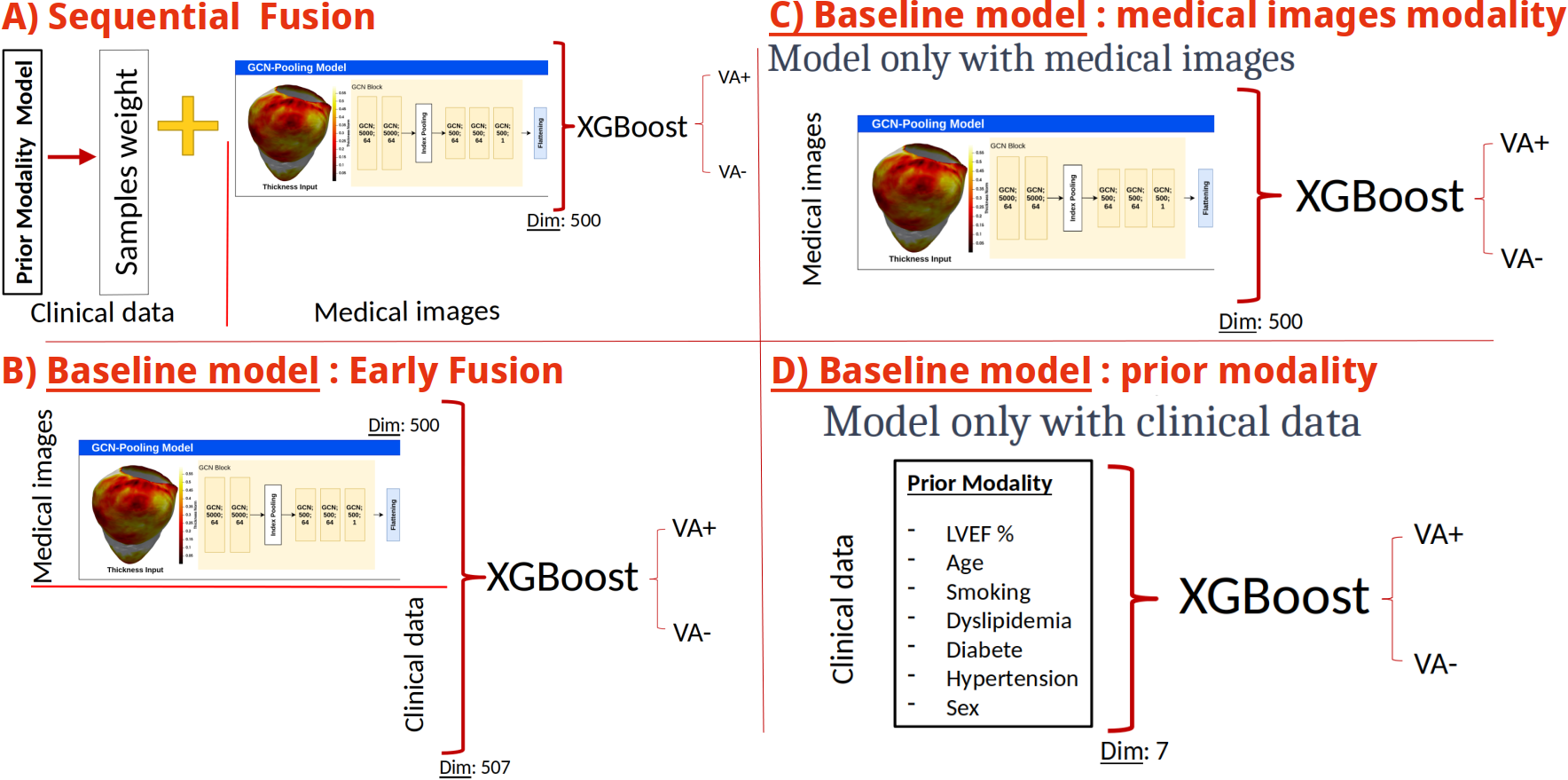}
  \caption{Different model configurations compared.}
  \label{fig:models_setting}
\end{figure}

The Early Fusion model (see fig. \ref{fig:models_setting}) is the concatenation of extracted thickness features with patients’ clinical data. The GCN-Pooling Model is used to extract thickness features. The data tensor is flattened and then given as input to a XGBoost classifier. The Early Fusion is the baseline model with both modalities.


The Sequential Fusion and Early Fusion models are compared with two others models used as a baseline with single modality. The first single modality baseline model is the prior model train with XGBoost classifier and only with clinical data (prior modality). The second single modality baseline model is the GCN-Pooling Model \cite{buntheng_ref} where the “FCN block” is replaced by an XGBoost classifier; this model is trained only with the medical images modality.
We use the XGBoost model implemented in the xgboost python package \cite{Tianqi_xgboost} as the base classifier. 
Same hyper-parameters are used for the four models with the number of estimators set to $300$ and the maximum depth is set to $6$. 
The default values, as proposed by the \cite{Tianqi_xgboost} package are kept for the other hyper-parameters. 

To isolate the impact of multimodal learning, we employed the XGBoost classifier across our model settings, acknowledging that the architecture of deep learning models (such as "FCN block") can vary depending on the input data. This approach ensures that any observed differences in performance are attributable to the multimodal learning techniques rather than variations in classifier architecture. We chose to utilise the TM rather than raw CT images for the VA classifier, as myocardial scar characterisation—derived from the TM—is a well-recognised substrate for VA \cite{Noordman_va_subtrate} \cite{Pandozi_va_substrate}.

\subsection{Controlled cross-validation setup}

In the realm of machine learning, the quality and structure of the training data significantly impact the performance and generalisation of a machine learning model. In our context, in addition to keeping the same ratio VA- over VA+ in each split, we must also check that each split (training, validation and testing set) has the same data distribution as the whole data set (used as reference). We use the T-test (uni-variate hypothesis testing on each patients' clinical data) to only keep triple splits (training, validation and testing set) which have the same distribution as our reference. The $p$-value significance threshold for the T-tests was set to $0.05$.

To compare the four models in our setting we have selected Sensitivity and F1 Score for performance metrics. 
To measure the variability of our performance metrics, 10 cross-validations are performed, then we compute the mean and standard deviation of each metric.

\section{Results}

As shown in Table.~\ref{models_perf_without_feat_5}, the Sequential Fusion model outperforms the Early Fusion model in term of both metrics F1 Score (+10\%) and Sensitivity (+15\%). On the F1 Score, it performs 21\% better than the prior modality model and 12\% better than the Medical images modality model. Looking at the Sensitivity metric, the  Sequential Fusion performs 18\% better than the prior modality model and 18\% better than the Medical images modality model.

\begin{table}[h]
\centering
\caption{Model performance results. Best performance metrics in \bf{bold}.}
\label{models_perf_without_feat_5}
    \begin{tabular}{l>{\centering\arraybackslash}p{2cm}>{\centering\arraybackslash}p{2cm}>{\centering\arraybackslash}p{2cm}>{\centering\arraybackslash}p{2cm}>{\centering\arraybackslash}p{2cm}}
    \toprule
        Model & Accuracy & F1 Score & Sensitivity & Specificity\\
        \midrule
        {(A) Sequential Fusion} & \bf 0.833: $\pm0.04$ & \bf 0.731: $\pm0.06$ & \bf 0.807: $\pm0.04$ & \bf 0.844: $\pm0.06$ \\
        \hline
        {(B) Early Fusion} & 0.790: $\pm0.03$ & 0.631: $\pm0.06$  & 0.650: $\pm0.08$ & \bf 0.844: $\pm0.04$ \\
        \hline
        {(C) Medical Images Modality} & 0.773:$\pm0.04$ & 0.607: $\pm0.05$  & 0.629: $\pm.06$ & 0.828: $\pm0.05$ \\
        \hline
        {(D) Prior modality} & 0.682: $\pm0.04$ & 0.521: $\pm0.06$  & 0.628: $\pm0.10$ & 0.703: $\pm0.05$ \\
        \midrule
    \end{tabular}
\end{table}

\section{Discussion}
\begin{figure}[!ht]
    \centering
    \begin{subfigure}[b]{0.57\textwidth}
        \centering
        \includegraphics[width=\textwidth]{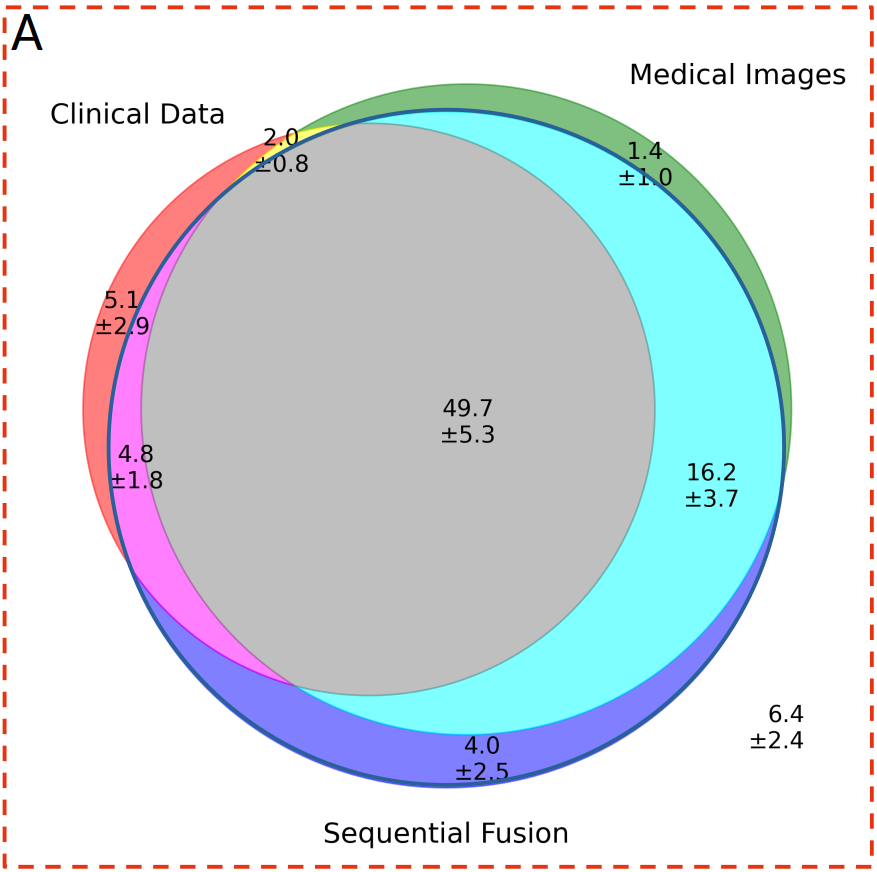}
        \label{fig:venn_seq_modalities}
    \end{subfigure}
    \hfill
    \begin{subfigure}[b]{0.39\textwidth}
        \centering
        \includegraphics[width=\textwidth]{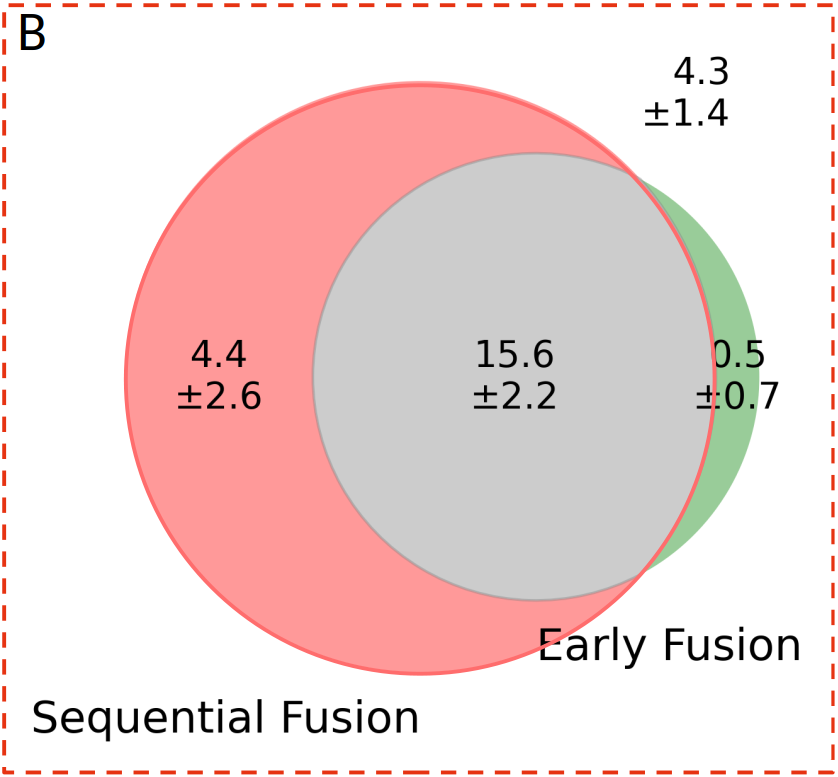}
        \label{fig:venn_seq_and_early}
    \end{subfigure}
    \caption{\textbf{A}: Venn diagram comparing well-classified samples by each modalities and Sequential Fusion model. Values are means and standard deviations on tests set. The clinical data classifier is red, the medical images classifier is green, and the Sequential Fusion classifier is blue. Overlapping regions are samples well-classified by corresponding overlapping classifiers. Overlapping regions colour are the mix of corresponding colours red, green or blue, except the intersection of all models which is in gray. Out of circles is the statistic of misclassified samples by all models.
    \textbf{B}: Venn diagram comparing well-classified positive samples (VA+) by Early Fusion and Sequential Fusion models. Sequential Fusion classifier is in light red, Early Fusion in light green and the intersection in gray.}
    \label{fig:venn_diagrams}
\end{figure}

\FloatBarrier
The objective of modalities fusion is to get the best from each modality according to the purpose of interest. When we look at the details, with fig. \ref{fig:venn_diagrams}-A, which shows samples well-classified by each modality (statistics on tests set), we can see that the Sequential Fusion is able to capture almost all knowledge from medical images modality. However, some samples well-classified by the prior modality are still misclassified by the Sequential Fusion. This point highlights the fact that there is still a room of progress to fuse modalities.
When comparing the classification results of Early Fusion and Sequential Fusion on positive samples (VA+), we observe an interesting point: Sequential Fusion captures nearly all the correctly classified VA+ samples identified by Early Fusion, as shown in fig. \ref{fig:venn_diagrams}-B.
In addition, we can also observed that some samples are misclassified by all models. Those samples may required another modality to improve the Sequential Fusion model.

\begin{figure}[!ht]
    \centering
    \begin{subfigure}[b]{0.55\textwidth}
        \centering
        \includegraphics[width=\textwidth]{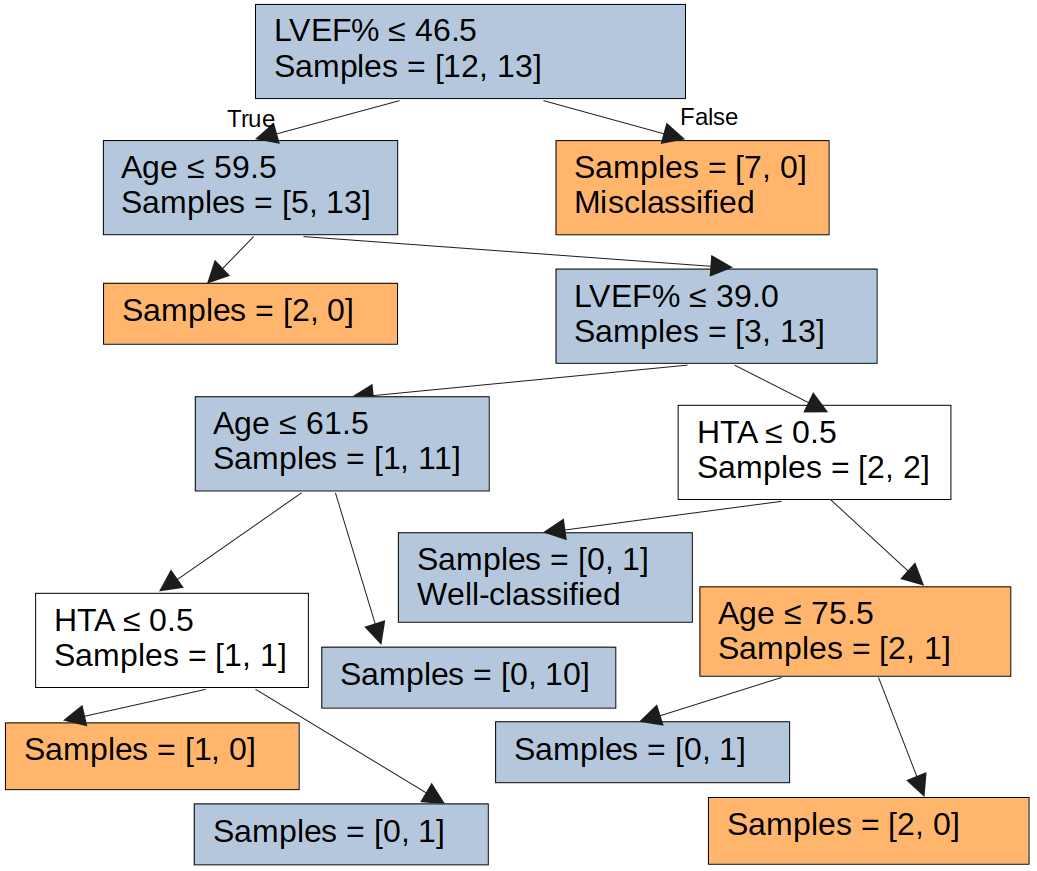}
        \caption{\textbf{Prior Modality}}
        \label{fig:prior_decision}
    \end{subfigure}
    \hfill
    \begin{subfigure}[b]{0.35\textwidth}
        \centering
        \includegraphics[width=\textwidth]{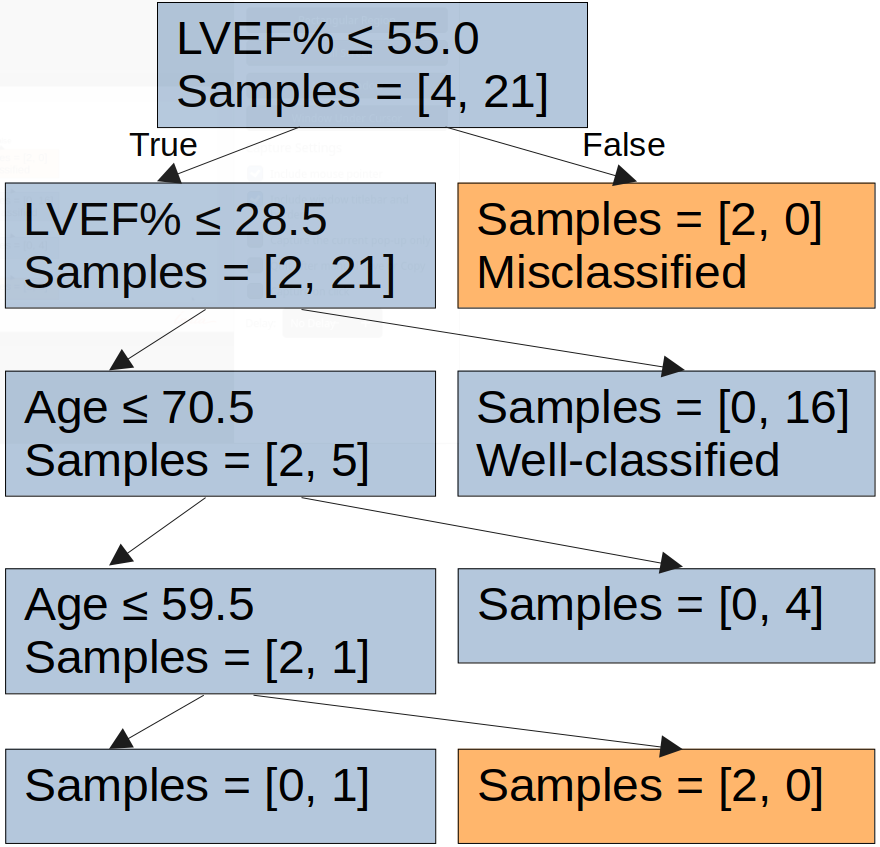}
        \caption{\textbf{Sequential Fusion}}
        \label{fig:seq_decision}
    \end{subfigure}
    \caption{Classification decision tree, of positive samples (VA+), on test set.}
    \label{fig:decision_trees}
\end{figure}

Fig. \ref{fig:decision_trees} gives an insight into the contribution of the medical images modality to the classification of VA+ samples, by comparing the decision tree of prior modality and Sequential Fusion. In fact, we can notice that the medical images modality has improved the classification of positive samples (VA+) with Left Ventricular (LV) Ejection Fraction greater than 46.5\%. This point is in agreement with the fact that cardiac imaging techniques have allowed improved SCD risk stratification, especially in the group of samples with an LVEF > 35\% \cite{van_der_scd_risk}.


\FloatBarrier
\section{Conclusion}

We have presented a novel technique to fuse two modalities by sequential learning on each modality. The main novelty of this technique is the weights computation based on knowledge acquired from the prior modality.
Our Sequential Fusion outperforms the Early Fusion with the ability to get most information from both modalities. 
When we look at misclassified samples by the Sequential Fusion and misclassified samples by all modalities, there is potential for growth, on the one hand, for fusion techniques, and for the inclusion of additional modalities on the other hand.
For future work on features fusion techniques, we believe that constraining progressively higher dimensional data based on lower dimensional data can improve the robustness of the classification model.

\section{\ackname} This work has been supported by the French government through France 2030, the National Research Agency (ANR) Investments in the Future with 3IA Côte d'Azur (ANR-19-P3IA-000) and LIRYC (ANR-10-IAHU-04).

%
%
%
%

\end{document}